\documentclass[iop]{emulateapj}

\usepackage{apjfonts}
\usepackage{float}
\usepackage{times}
\usepackage{natbib}
\usepackage{rotating}
\makeatother 

\lefthead{Geach et al.} \righthead{A cold, clumpy, illuminated outflow around
SSA22-LAB1}

\begin{document}

\title{A submillimeter galaxy illuminating its
circumgalactic medium: Ly$\alpha$ scattering in a cold, clumpy outflow}

\author{J.~E.~Geach\altaffilmark{1},
R.\ G.\ Bower\altaffilmark{2}, 
D.\ M.\ Alexander\altaffilmark{2}, 
A.\ W.\ Blain\altaffilmark{3},
M.\ N.\  Bremer\altaffilmark{4}, 
E.\ L.\ Chapin\altaffilmark{5},
S.\ C.\ Chapman\altaffilmark{6},
D.\ L.\ Clements\altaffilmark{7},
K.\ E.\ K.\ Coppin\altaffilmark{1},
J.\ S.\ Dunlop\altaffilmark{8},
D.\ Farrah\altaffilmark{9},
T.\ Jenness\altaffilmark{10,11}, 
M.\ P.\ Koprowski\altaffilmark{8},
M.\ J.\ Micha\l owski\altaffilmark{8},
E.\ I.\ Robson\altaffilmark{12}, 
D.\ Scott\altaffilmark{13}, 
D.\ J.\ B.\ Smith\altaffilmark{1}, 
M.\ Spaans\altaffilmark{14},
A.\ M.\ Swinbank\altaffilmark{2},
P.\ van der Werf\altaffilmark{15}}
\altaffiltext{1}{Center for Astrophysics Research, Science \& Technology Research Institute, University of Hertfordshire, Hatfield, AL10 9AB, UK. j.geach@herts.ac.uk}
\altaffiltext{2}{Institute for Computational Cosmology, Department of Physics, Durham University, South Road, Durham, DH1 3LE, UK}
\altaffiltext{3}{Department of Physics \& Astronomy, University of Leicester, University Road, Leicester, LE1 7RH, UK}
\altaffiltext{4}{School of Physics, HH Wills Physics Laboratory, Tyndall Avenue, Bristol BS8 1TL, UK}
\altaffiltext{5}{XMM SOC, ESAC, Apartado 78, 28691 Villanueva de la Canada, Madrid, Spain}
\altaffiltext{6}{Department of Physics and Atmospheric Science, Dalhousie University Halifax, NS, B3H 3J5, Canada}
\altaffiltext{7}{Astrophysics Group, Imperial College London, Blackett Laboratory, Prince Consort Road, London, SW7 2AZ, UK}
\altaffiltext{8}{Institute for Astronomy, University of Edinburgh, Royal Observatory, Blackford Hill, Edinburgh EH9 3HJ, UK}
\altaffiltext{9}{Virginia Polytechnic Institute \& State University  Department of Physics, MC 0435, 910 Drillfield Drive, Blacksburg, VA 24061, USA}
\altaffiltext{10}{Joint Astronomy Centre 660 N. A'ohoku Place University Park Hilo, Hawaii 96720, USA}
\altaffiltext{11}{Department of Astronomy, Cornell University, Ithaca, NY 14853, USA}
\altaffiltext{12}{UK Astronomy Technology Centre, Royal Observatory, Blackford Hill, Edinburgh EH9 3HJ, UK}
\altaffiltext{13}{Department of Physics \& Astronomy, University of British Columbia, 6224 Agricultural Road, Vancouver, BC, V6T 1Z1, Canada}
\altaffiltext{14}{Kapteyn Institute, University of Groningen, P.O. Box 800, 9700 AV Groningen, The Netherlands}
\altaffiltext{15}{Leiden Observatory, Leiden University, P.O. box 9513, 2300 RA Leiden, The Netherlands}

%\maketitle 

\label{firstpage}

\begin{abstract}We report the detection at 850\,$\mu$m of the central source
in SSA22-LAB1, the archetypal `Lyman-$\alpha$ Blob' (LAB), a 100\,kpc-scale
radio-quiet emission-line nebula at $z=3.1$. The flux density of the source,
$S_{\rm 850}=4.6\pm1.1$\,mJy implies the presence of a galaxy, or group of
galaxies, with a total luminosity of $L_{\rm IR}\approx10^{12}\,{\rm
L_\odot}$. The position of an active source at the center of a
$\sim$50\,kpc-radius ring of linearly polarized Ly$\alpha$ emission detected
by Hayes et al.\ (2011) suggests that the central source is leaking Ly$\alpha$
photons preferentially in the plane of the sky, which undergo scattering in
H{\sc i} clouds at large galactocentric radius. The Ly$\alpha$ morphology
around the submillimeter detection is reminiscent of biconical outflow, and
the average Ly$\alpha$ line profiles of the two `lobes' are dominated by a red
peak, expected for a resonant line emerging from a medium with a bulk velocity
gradient that is outflowing relative to the line center. Taken together, these
observations provide compelling evidence that the central active galaxy (or
galaxies) is responsible for a large fraction of the extended Ly$\alpha$
emission and morphology. Less clear is the history of the cold gas in the
circumgalactic medium being traced by Ly$\alpha$: is it mainly pristine
material accreting into the halo that has not yet been processed through an
interstellar medium (ISM), now being blown back as it encounters an outflow,
or does it mainly comprise gas that has been swept-up within the ISM and
expelled from the galaxy?

\end{abstract}

\keywords{galaxies: active, galaxies: formation, galaxies: high-redshift}
\section{Introduction: Lyman-$\alpha$ Blobs}

Lyman-$\alpha$ Blobs (LABs) are $\sim$100\,kpc scale radio-quiet
Lyman-$\alpha$ emission line nebulae, with integrated luminosities up to
$L_{\rm Ly\alpha}\sim10^{37}$\,W. LABs were first discovered during the course
of deep optical narrowband imaging of the Small Selected Area 22$^{\rm hr}$
(SSA22) field (Steidel et al.\ 2000), which was known to contain a significant
overdensity of Lyman-break Galaxies (LBGs) at $z=3.09$ (Steidel et al.\ 1998),
although previously Francis et al.\ (1996) and Keel et al.\ (1999) had also
detected extended Ly$\alpha$ emission around luminous galaxies at $z=2.4$
(both in high density environments) and extended Ly$\alpha$ emission was
reported to be associated with some of the first high-{\it z} submillimetre
galaxies (SMGs, e.g., Ivison et al.\ 1998). Subsequent narrowband (and
broadband, see Prescott et al.\ 2012, 2013) surveys of the SSA22
`protocluster' and elsewhere have since revealed a substantial sample of these
high redshift nebulae, with a range of Ly$\alpha$ luminosities and extents
(e.g.,\ Matsuda et al.\ 2004, 2011, Yang et al.\ 2009, Erb et al.\ 2011, Bridge
et al.\ 2013) and often inhabiting dense environments (e.g.\ Prescott et
al.\ 2008).

Since their discovery, the nature of LABs has been in debate. It was
recognised early on that the galaxy halo cooling budget would contain a
significant cold component (White \& Frenk\ 1991), and low metallicity gas at
$T\sim10^4$--$10^5$\,K cools most efficiently via Ly$\alpha$ emission through
collisional excitation (Fabian \& Nulsen\ 1977, Katz \& Gunn\ 1991, Katz et
al.\ 1996, Haiman et al.\ 2000), so one train of thought is that extended
Ly$\alpha$ emission implies the presence of potentially pristine cold gas in
the circumgalactic medium (CGM). Taking this idea further, many cosmological
hydrodynamical simulations suggest that the baryonic growth of massive
($M_{\rm h}>10^{12}\,{\rm M_\odot}$) galaxies at $z>2$ (and for lower mass
galaxies at {\it all} epochs) is dominated by a mode of `cold accretion' in
the form of narrow filamentary streams that penetrate the hot, virially
shocked halo gas (Katz et al.\ 2003, Kere\v{s} et al.\ 2005,\ 2009, Dekel et
al.\ 2009). Nevertheless, the predicted cold flows have so-far eluded
unambiguous observation and some recent simulations seem to refute the
existence of cold flows that survive to the disc altogether (Nelson et al.\
2013). Nevertheless, LABs have been touted as the closest we have come to
identifying the filamentary cold mode accretion phenomenon in nature.

While the cosmological gas simulations can track the history of cold gas
easily enough, the predicted fluxes of the cooling Ly$\alpha$ emission are far
more uncertain, with the flux depending on the complex interplay of
turbulence, radiative transfer and local sources of ionizing radiation are
taken into account (e.g.,\ Rosdahl \& Blaziot\ 2012). The prescriptions for
Ly$\alpha$ radiative transfer through the gas are notoriously complex due to
the fact that it is a resonant line (Neufeld\ 1990) and sensitive to
assumptions about the sub-grid physics governing the transport of Ly$\alpha$
photons through astrophysical media, as well as the many other possible
sources of Ly$\alpha$ photons in and around galaxies. As a result, different
simulations that apply bespoke models of Ly$\alpha$ radiative transfer, aiming
to reproduce the properties of observed LABs, predict a wide range of physical
properties (Ly$\alpha$ luminosities can vary by an order of magnitude) and key
observables, such as Ly$\alpha$ extents and surface brightnesses
(Faucher-Gigu\`ere et al.\ 2010, Goerdt et al.\ 2010, Rosdahl \& Blaziot
2012).

The cold flow picture is complicated further when one considers the effects of
the presence of a luminous source (a starburst galaxy or one hosting an active
galactic nucleus, AGN) embedded within the LAB, which appears to be a common
situation (e.g.,\ Francis et al.\ 1996, Keel et al.\ 1999, Dey et al.\ 2005,
Geach et al.\ 2005,\ 2007,\ 2009, Colbert et al.\ 2006, Webb 2009, Rauch
et al.\ 2011, Cantalupo, Lilly \& Haenhelt\ 2012). The typical bolometric
luminosities of the embedded sources imply an energy balance heavily weighted
towards the galaxy, such that only relatively weak coupling, through feedback,
would be required to explain the Ly$\alpha$ luminosities (Geach et al.\ 2009).
Again, there is further uncertainty in the type of feedback that would give
rise to such extended Ly$\alpha$ emission: proposals have included emergent
(supernovae-driven) superwinds that shock heat cold gas swept up in the
outflow (Taniguchi \& Shioya\ 2000), photoionization of the CGM by escaping
ultraviolet continuum radiation from massive stars or the accretion discs of
an AGN, or photoionization via inverse Compton (IC) scattering of cosmic
microwave background (or locally generated far-infrared) photons, which are
energetically up-scattered by a relic population of relativistic electrons
(e.g.,\ Scharf et al.\ 2003, Fabian et al.\ 2009).

Geach et al.\ (2009) rule-out IC scattering as a viable LAB formation
mechanism using a deep X-ray stack in SSA22, arguing that photoionization by a
central AGN (which may be present in up to 50\% of LABs) is probably more
important, even with fairly modest escape fractions likely to be applicable to
the often heavily obscured sources. Recently, Hayes et al.\ (2011) presented
evidence that at least some of the extended Ly$\alpha$ emission must be in the
form of scattered Ly$\alpha$ radiation, with Ly$\alpha$ photons generated
within a central galaxy scattering in clouds of H{\sc i} at large
($\sim$50\,kpc) galactocentric radius. The absence of an obvious luminous
counterpart in some LABs (e.g.,\ Nilsson et al.\ 2006, Smith \& Jarvis\ 2007,
Smith et al.\ 2008) has been inferred to be a sign that, in some cases at
least, cooling radiation alone is sufficient to explain the extended
Ly$\alpha$ emission. However, often the constraints on the level of AGN or
starburst activity are too poor to rule out the involvement of feedback energy
completely.

In practice, it seems likely that a mixture of gravitational cooling and
feedback processes are at play in LABs, each imparting an ambiguous
observational signature in the Ly$\alpha$ emission that is further muddied by
observational limitations (namely angular and spectral resolution, and poor
sensitivity to low surface-brightness features). In any case, it is important
to note that {\it both} the cooling and `heating' models require a reasonable
reservoir of gas in the circumgalactic medium, so whatever the process
dominating the Ly$\alpha$ emission, LABs provide a rare opportunity to
investigate the astrophysics at the galaxy/intergalactic medium interface
close to the formation epoch of massive galaxies, $z\approx2$--$3$. 

\section{SSA22-LAB1}

The archetypal giant LAB, SSA22-LAB1, is the most scrutinized specimen of this
class of object, but it is perhaps also the most controversial. Chapman et al.\ (2001,
2004) reported the detection of a very bright submillimeter galaxy at the
center of the nebula, with a 850\,$\mu$m (James Clerk Maxwell Telescope
[JCMT]/SCUBA) flux of $\sim$17\,mJy, implying a galaxy with hyperluminous
levels of activity, which Geach et al.\ (2007) identified with a {\it
Spitzer}-IRAC source likely to be the counterpart. However it appears this
measurement over-estimated the submillimeter emission. Matsuda et al.\ (2007)
observed SSA22-LAB1 at higher resolution ($2''$) using the SubMillimeter Array
(SMA), at a 1$\sigma$ depth of 1.4\,mJy\,beam$^{-1}$ at 880\,$\mu$m. The target
was not detected, and the conclusion was that the 850\,$\mu$m emission reported
by Chapman et al.\ must be extended on scales of several arcseconds, and
resolved out in the SMA map. The mystery deepened when Yang et al.\ (2012)
reported no detection of SSA22-LAB1 at 870\,$\mu$m in an APEX/LABOCA
observation, at a similar resolution to the original SCUBA observation (using
the on/off chopping technique in photometry mode). The reported 3$\sigma$
upper limit of $S_{\rm 870}<12$\,mJy is clearly at odds with the SCUBA
detection. Most recently, in an AzTEC/ASTE 1.1\,mm map of the field, Tamura et
al.\ (2013) report only a marginally significant 2.7$\sigma$ (1.9\,mJy) flux
enhancement at the position of SSA22-LAB1, again in tension with the original
SCUBA observations.

In this paper we present new JCMT/SCUBA-2 (Holland et al.\ 2013) observations
of SSA22-LAB1, taken as part of the SCUBA-2 Cosmology Legacy Survey (Geach et
al.\ 2013, Roseboom et al.\ 2013) and interpret the results in the context of
extremely deep Ly$\alpha$ SAURON integral-field observations (Weijmans et al.\
2010) and the recent Very Large Telescope/UV FOcal Reducer and low dispersion
Spectrograph (FORS) Ly$\alpha$ polarimetry observations of Hayes et al.\
(2011). The remainder of the paper is organised as follows: in \S3 we describe
the SCUBA--2 observations, \S4 presents the main result and we interpret and
discuss this in context with other observations in \S5. We conclude the paper
with a hypothesis for the origin of SSA22-LAB1, and propose further questions
in \S6. A $\Lambda$CDM cosmology defined by the parameters measured with the
{\it Wilkinson Microwave Anisotropy Probe} (7 year results including baryonic
acoustic oscillation and Hubble constant constraints, Komatsu et al.\ 2011) is
assumed throughout: $h=0.70$, $\Omega_{\rm m}=0.27$, $\Omega_\Lambda=0.73$.

\begin{figure*}
\centerline{\includegraphics[height=3.1in]{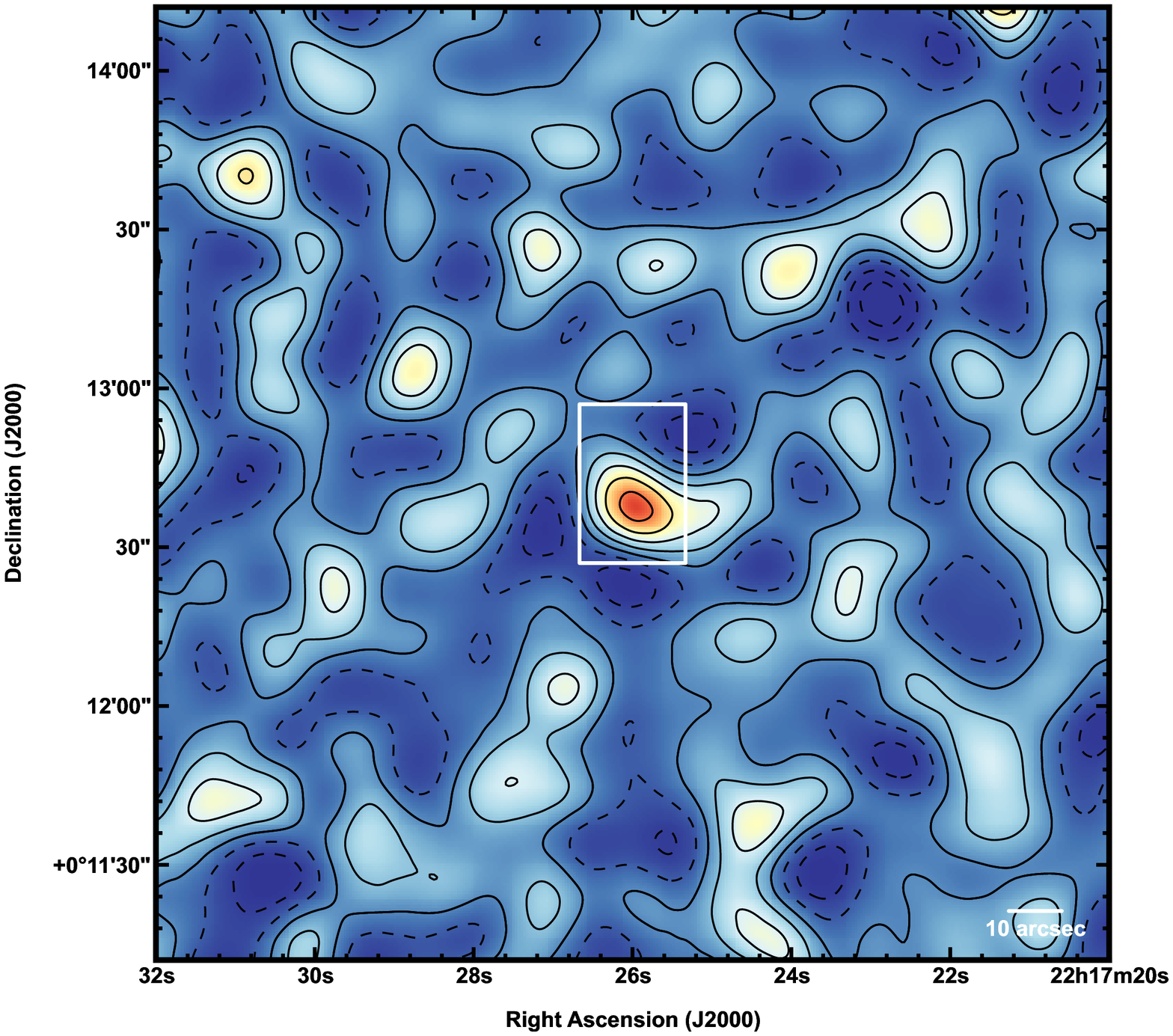}\includegraphics[height=3.1in]{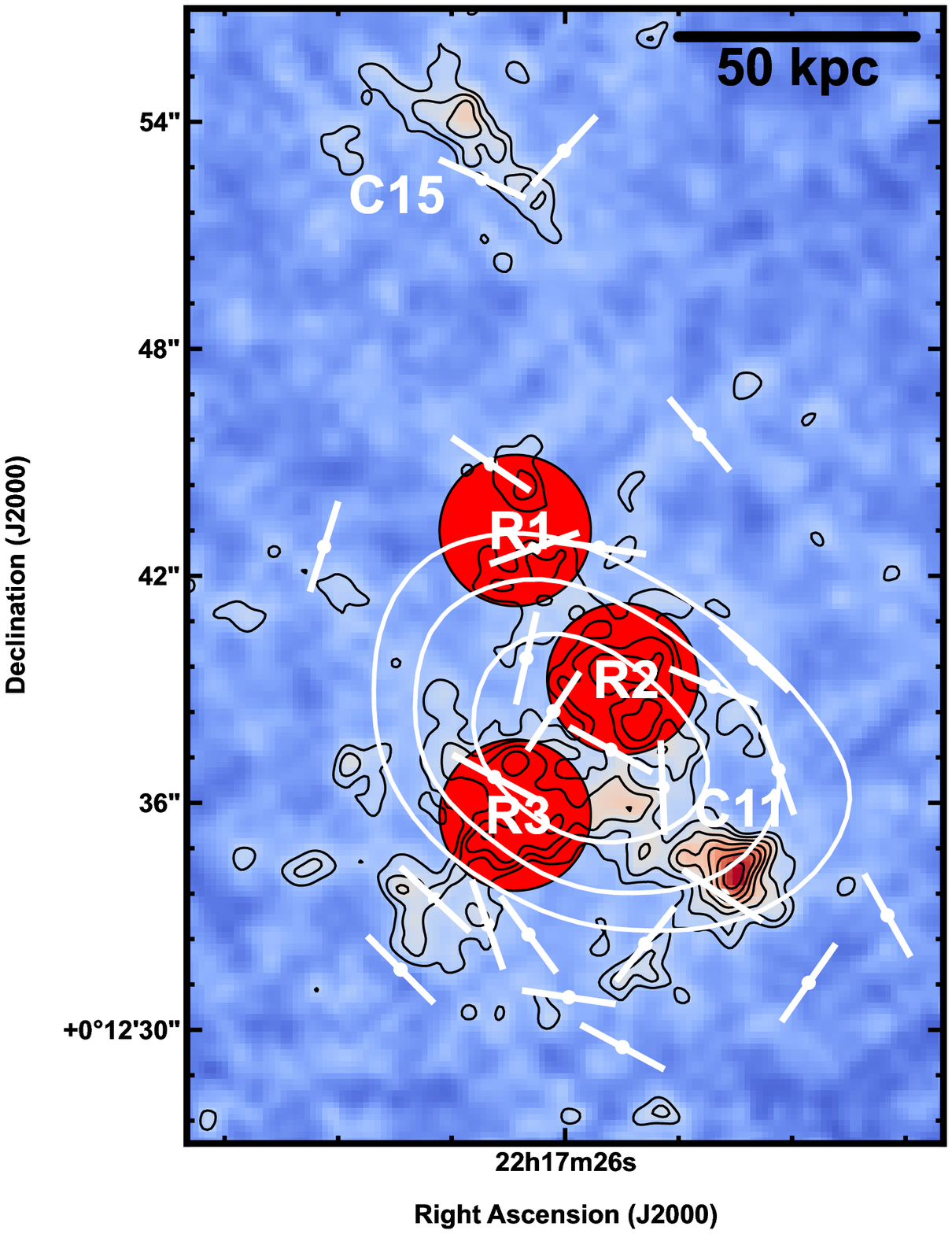}}
\caption{(Left) SCUBA--2 850\,$\mu$m map of the region $(3'\times3')$ around
SSA22-LAB1; contours indicate significance levels of $-5\leq\sigma\leq5$ in
steps of 1$\sigma$, where $1\sigma=1.1$\,mJy\,beam$^{-1}$. We detect a source
with $S_{850}=(4.6\pm 1.1)$\,mJy\,beam$^{-1}$ at the position of SSA22-LAB1,
and confirm that a $\sim$3$\sigma$ detection is measured at the same position
in independent halves of the data (\S4). The white box shows the SAURON field
of view, detailed in the second panel. (Right) zoom-in of SSA22-LAB1 showing
the average line flux over 4922--5022\AA\ observed in a 24\,hour integration
with the SAURON IFU (Matsuda et al.\ 2004, Weijmans et al.\ 2010). We show the
SCUBA-2 detection as large white contours (0.5$\sigma$ levels starting at
3$\sigma$). The LAB clearly breaks up in to distinct regions, two of which are
centered on Lyman-break galaxies (C11 and C15) at $z=3.1$, whereas R1-R3 are
more diffuse and not obviously connected with an ultraviolet continuum source.
White lines show the direction of the polarization pseudovectors
($\chi=0.5\arctan(U/Q)$, where $U$ and $Q$ are the Stokes parameters, Hayes et
al.\ 2011). The polarization pattern describes a ring surrounding the SMG, and
we postulate that this is the source of Ly$\alpha$ photons, which are escaping
the galaxy and scattering in H{\sc i} clouds at large radius.} \end{figure*}

\section{SCUBA--2 observations}

The SSA22 field has been observed as part of the JCMT SCUBA--2 Cosmology
Legacy Survey (S2CLS). A 30 arcminute diameter map, centered on 22$^{\rm
h}$17$^{\rm m}$36.3$^{\rm s}$, $+$00$^\circ$19$'$22.7$''$, has been obtained
(using multiple repeats of the PONG scanning pattern). A total of 105
observations were made between August 23, 2012 and November 29, 2013 in
conditions when the zenith optical depth at 225\,GHz was in the range
$0.05<\tau_{225}<0.1$, with a mean of $\langle \tau_{225} \rangle=0.07$. The
beam-convolved map reaches a 1$\sigma$ depth of 1.1\,mJy\,beam$^{-1}$ with an
integration time of $\sim$3000\,seconds per 4 arcsecond pixel. Data are
reduced using the {\sc smurf} {\it makemap} pipeline (Chapin et al.\ 2013),
following the procedures described in more detail in Geach et al.\ (2013).
Flux calibration is performed using the best estimate of the flux conversion
factor based on observations of hundreds of standard calibrators (Dempsey et
al.\ 2013); this absolute flux calibration is expected to be accurate at the
15\% level.

\section{Results}

A submillimeter source with $S_{\rm 850}=(4.6\pm1.1)$\,mJy (error bar is
instrumental noise only) is detected at 22$^{\rm h}$17$^{\rm m}$26.0$^{\rm s}$, +00$^\circ$12$'$37.5$''$.
This is within 1.5$''$ of a red {\it Spitzer}-IRAC continuum counterpart
source ($S_{\rm 8}=7.6\pm2.2$\,$\mu$Jy, $S_{\rm 8}/S_{\rm 4.5}=1.3\pm2.2$),
close to the center of the Ly$\alpha$ halo that Geach et al.\ (2007)
identified as the central galaxy. To investigate the veracity of
the submillimeter detection, we create two `half maps' from the SCUBA-2 data,
whereby each half map represents a random 50\% of the individual scans used to
make the total map. At the sky position of the detection reported above, we
measure significant flux densities of $(4.8\pm1.6)$\,mJy and $(4.5\pm1.5)$\,mJy
(i.e.,\ 3$\sigma$ detections with consistent flux densities in independent
halves of the integration), which lend credence to the submillimeter detection
of SSA22-LAB1.

Chapman et al.\ (2004) also present a tentative VLA 1.4\,GHz radio continuum
source and Owens Valley Radio Observatory (OVRO) CO(4--3) detection consistent
with this position (although Yang et al.\ 2012 failed to detect CO(4--3) in
SSA22-LAB1 with IRAM PdBI). The SCUBA-2 map in the vicinity of SSA22-LAB1 is
shown in Fig.\ 1. The mid-infared colours of the IRAC counterpart suggest that
star formation dominates the energy budget of this source (Colbert et al.\
2011). Clearly the new 850\,$\mu$m flux density is lower than the $S_{\rm
850}=16.8\pm2.9$\,mJy reported by Chapman et al.\ (2001). Formally the
flux density ratio of the initial SCUBA measurement and the new SCUBA-2 flux
density is $3.7\pm1.1$; the origin of this disparity remains unclear.

The revised 850\,$\mu$m flux density is consistent with the LABOCA upper-limit
$S_{870}<12$\,mJy (Yang et al.\ 2012), and at the 1$\sigma$
1.4\,mJy\,beam$^{-1}$ limit of the higher-resolution ($2''$ beam) SMA map of
Matsuda et al.\ (2007), the 850\,$\mu$m emission would only have to be
distributed on a scale of a few arcseconds to fall below the SMA detection
rate. Still, the galaxy-integrated 850\,$\mu$m emission implies a vigorously
star-forming galaxy. With no other constraints on the shape of the
far-infrared spectral energy distribution, we can only make a rough estimate
of the bolometric (total infrared) luminosity. Assuming a nominal single
modified blackbody spectrum with $\beta=2$ and $T=20$--$30$\,K (Magnelli et
al.\ 2012), normalised to $S_{\rm 850}=4.6$\,mJy (the data do not allow a
constraint on either the emissivity or temperature), implies a $\log(L_{\rm
IR}/{\rm L_\odot})\approx 11.8$--$12.6$, where the infrared luminosity is
integrated over $\lambda=40$--$120$\,$\mu$m (rest frame) and includes a
bolometric correction term of 1.91 to account for hot dust emission at
mid-infrared wavelengths not modelled by the single blackbody (Helou et al.\
1988; Dale et al.\ 2001; Magnelli et al.\ 2012).

The integrated luminosity estimated above should be considered with a caveat:
the 15$''$ resolution of the SCUBA-2 map naturally results in a submillimeter
flux measurement that is integrated over the equivalent of $\sim$120\,kpc at
$z=3.1$, thus encompassing (potentially) several individual sources
contributing to the 850\,$\mu$m emission. It is already well known that a
significant fraction of submillimeter galaxies ($\approx$35\% and potentially
up to 50\%) detected with coarse resolution instruments break-up into several
components when observed at higher resolution, as is now possible with
sensitive interferometers (e.g.,\ Hodge et al.\ 2013). Given the negative {\it
k}-correction in the submillimeter bands on the Rayleigh-Jeans side of the
cool dust emission, chance alignments of luminous galaxies across a wide range
of epochs could give rise to multiplicity, however, given the likelihood that
many high-{\it z} submillimeter sources are either physically large, clumpy
systems or undergoing mergers means that single detections that resolve into
multiple components does not preclude them being considered part of the same
physical system. Obviously obtaining higher resolution submillimeter continuum
imaging of SSA22-LAB1 is now a goal, since that will address this question.

There are some hints that we might expect the SCUBA-2 850\,$\mu$m detection to
be extended: the failure to detect continuum emission in the 2$''$ resolution
SMA map of Matsuda et al.\ (2007) with a 3$\sigma$ upper limit of
4.2\,mJy\,beam$^{-1}$ suggests that the 850\,$\mu$m flux is distributed on a
scale of at least $>$2$''$. At the position of the IRAC counterpart,
SSA22-LAB1a identified by Geach et al.\ (2007), an {\it HST}-STIS image
reveals several distinct (but extremely faint) ultraviolet components spread
over several arcseconds (Chapman et al.\ 2004). For the sake of argument, in
the following we will refer to the central source as a single ULIRG-class
galaxy (since, even if the emission was distributed on scales of several
arcseconds, this is far smaller than the Ly$\alpha$ extent of the LAB), but
the reader should remember that at all times we are referring to the total
infrared luminosity sampled by the JCMT beam, and this could encompass several
distinct -- but physically associated -- sources.

\section{Interpretation and discussion}

The detection of a luminous galaxy in the core of SSA22-LAB1 is further
evidence that Lyman-$\alpha$ Blobs are generally associated with luminous
young galaxies, be it an obscured starburst or an AGN (Geach et al.\ 2009).
SSA22-LAB1 is unique in that it has been studied in far greater detail than
any other LAB, and therefore offers the best opportunity to understand the
astrophysics of the LAB phenomenon. Here we revisit the observations of
SSA22-LAB1 by Weijmans et al.\ (2010) and Hayes et al.\ (2011) and relate
these to the confirmation of a luminous central source.

\subsection{Spatial distribution and spectral properties of the Lyman-$\alpha$
emission}

Weijmans et al.\ (2010) present very deep WHT/SAURON integral field unit (IFU)
spectral observations of SSA22-LAB01 at $4810$\AA\,$<\lambda<$$5350$\AA,
combining the data with the previous observations of Bower et al.\ (2004) to
form a data cube with total integration time of nearly 24 hours. The IFU data
provide a much better probe of the structure of the Ly$\alpha$ halo than
narrow-band imaging alone (Steidel et al.\ 2000, Matsuda et al.\ 2004).

\begin{figure*}
\includegraphics[height=0.33\textwidth,angle=-90]{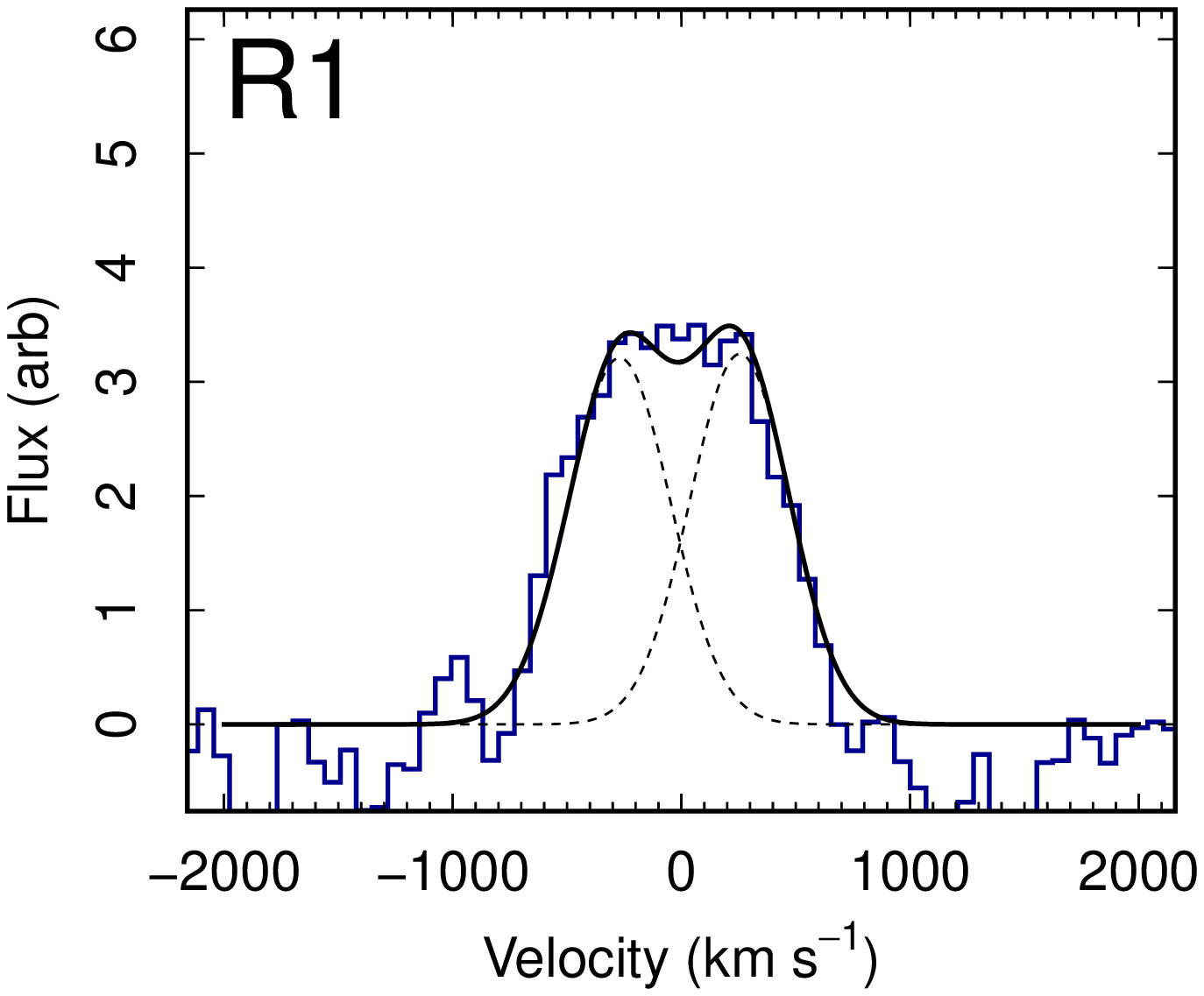}\includegraphics[height=0.33\textwidth,angle=-90]{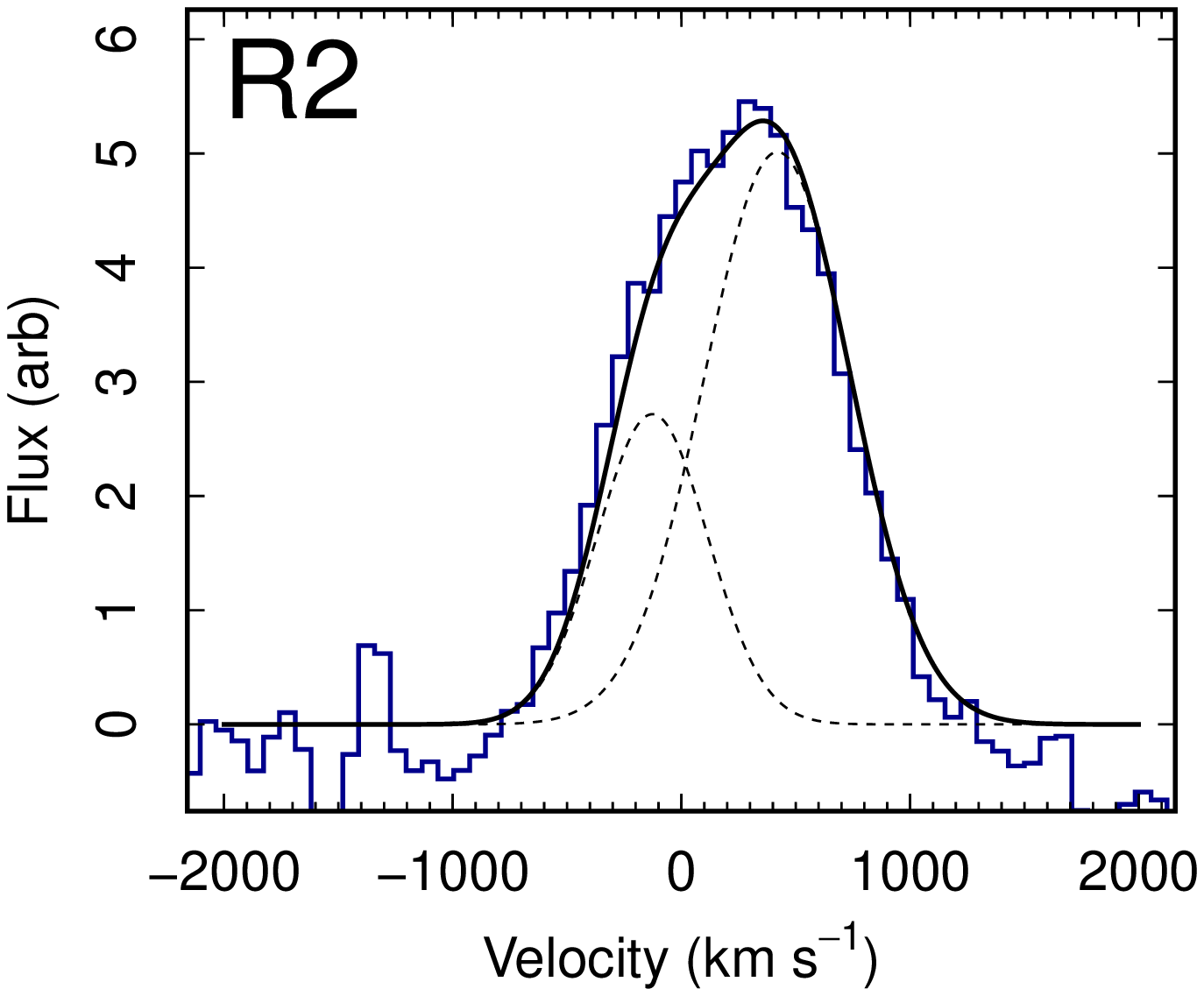}\includegraphics[height=0.33\textwidth,angle=-90]{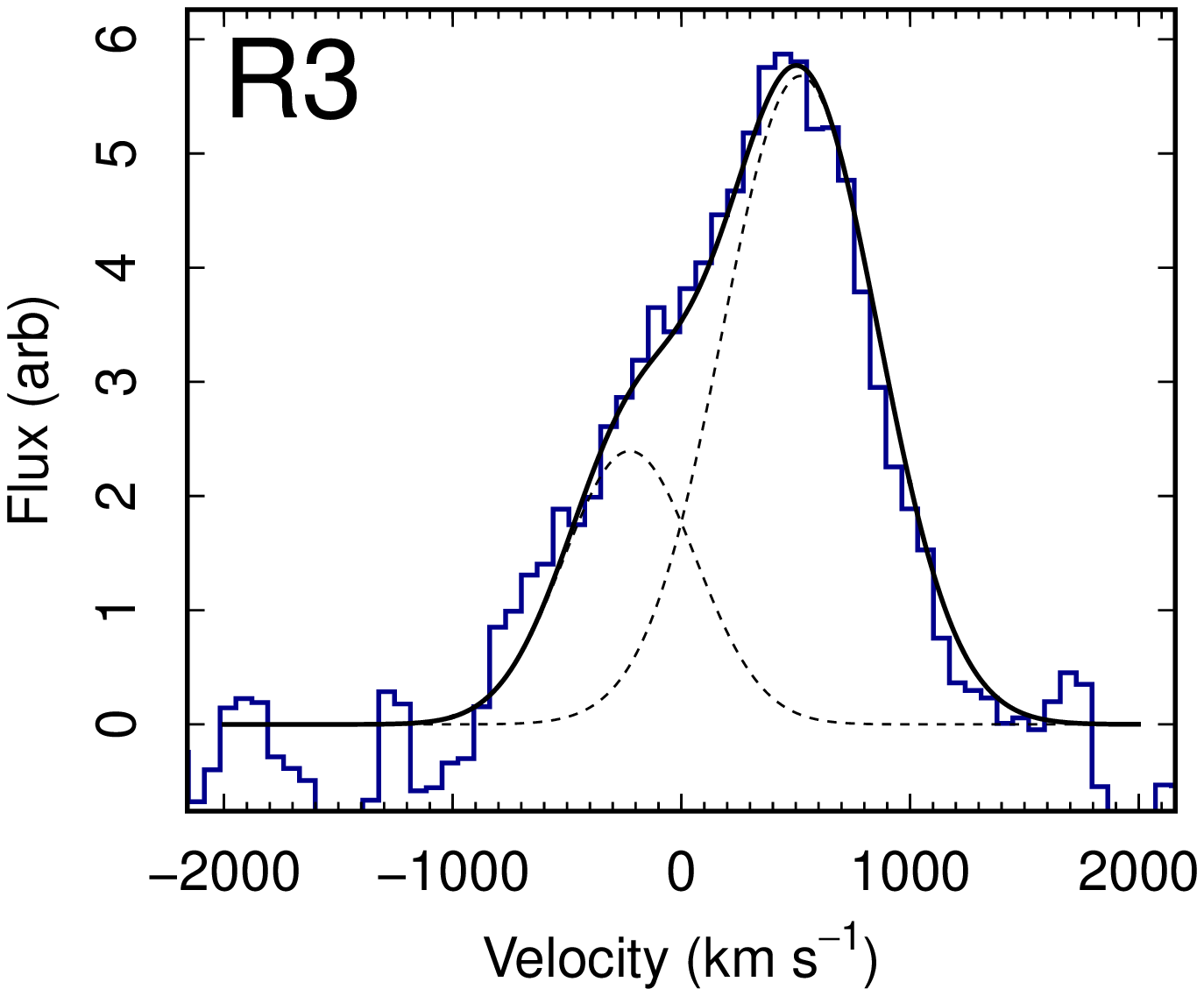}
\caption{Average Ly$\alpha$ line profiles extracted from the 4$''$-diameter
apertures R1--R3 indicated in Fig.\ 1, shown on a common, but arbitrary, flux
scale. Note that this figure is adapted from Weijmans et al.\ (2010), where we
have extracted average spectra from different apertures and independently fit
the data here. R1 is a low surface brightness structure extending north from
R2 towards the LBG C15, whereas the brighter R2 and R3 represent two `lobes'
of emission straddling the SMG. The lines are all broad and asymmetric, with
R2 and R3 showing prominent red peaks, expected if a bulk velocity gradient is
present that is positive relative to the line center, and usually interpreted
as outflowing material (we have taken into account the instrumental
dispersion, 108\,km\,s$^{-1}$). The profile of the R1 spur, in contrast, is
more symmetric, and it is unclear whether this gas is inflowing, outflowing or
static relative to the central galaxy.} \end{figure*}

Aside from Ly$\alpha$ emission associated with two Lyman Break Galaxies (C11
and C15), there are three main regions of diffuse Ly$\alpha$ emission,
labelled R1, R2 and R3 by Weijmans et al. These contain no ultraviolet
continuum source, but the two brightest components R2 and R3 ($L_{\rm
Ly\alpha}=2.3\times10^{36}$\,W and $2.7\times10^{36}$\,W respectively)
approximately straddle the peak of the 850\,$\mu$m emission that we have
identified with SCUBA-2. R1 is best described as a lower surface brightness
stucture extending north from R2 towards C15. Is there a physical
motivation for associating the diffuse Ly$\alpha$ emission with the central
submillimetre source? In Fig.\ 2 (adapted from Weijmans et al.\ 2010) we
present the average Ly$\alpha$ spectra extracted from 2$''$ radius apertures
at the positions R1, R2 and R3 indicated on Fig.\ 1. All the lines are broad,
with full width zero intensity widths up to $\sim$2000\,km\,s$^{-1}$, and
poorly fit by single Gaussian dispersion profiles. As Weijmans et al. show, a
double Gaussian fit is a better model of the line profiles in all three cases
(a single Gaussian combined with a Voigt absorption profile also provides an
adequate fit to the data). In Fig.\ 2 we show the best fitting double Gaussian
profiles, centered in the rest frame of each component. R2 and R3 are within
180\,km\,s$^{-1}$ of each other, and both profiles have a dominant red-skewed
peak. In contrast, R1 has a slightly more symmetric profile and is separated
from R2 by approximately 300\,km\,s$^{-1}$.

For resonant line radiation transfer through an optically thick slab the
emergent line profile is symmetric and double peaked, with a minimum at
$\Delta v=0$ due to scattering at the line resonance (Neufeld\ 1990). In an
H{\sc i} cloud, the separation of the peaks is set by the gas temperature and
column density, which determine the typical frequency change a photon can
experience through scattering. If a large-scale velocity gradient is present,
then the line profile becomes asymmetric, with either the red or blue peak
dominating, depending on the direction of the velocity gradient with respect
to the resonant frequency: outflows generally result in a prominent red peak,
whereas infall results in a dominant blue peak ($z<z_{\rm em}$ Ly$\alpha$
Forest absorption suppresses the blue wing whatever the situation). Turbulent
motions within the medium will also broaden the emergent line, and asymmetric
line profiles can also occur due to foreground absorption of Ly$\alpha$
photons by clumps, or screens, of dust or H{\sc i} (as is proposed for the
line profile observed in LAB2, see Wilman et al.\ 2005). 

It is interesting that the line profiles of R2 and R3 both have dominant red
peaks. Given their spatial distribution relative to the SMG, one could
interpret this as evidence of a biconical outflow driven by the central ULIRG.
Photons from the central source that are back-scattered from the interior,
rear wall of the expanding shell will be redshifted out of resonance with the
front wall, while photons forward scattered from the front wall remain in
resonance with the expanding shell. Note that the bipolar appearance does not
necessarily require that the outflow is bipolar: the emergent structure could
be equally well explained if ionizing photons only escaped from the central
source along preferential directions.

\subsection{Polarized Lyman-$\alpha$ emission: scattering of photons from a
central ULIRG?}

Hayes et al.\ (2011) used VLT/FORS to measure the polarization of the extended
Ly$\alpha$ emission in SSA22-LAB1. In Fig.\ 1 we overlay the polarization
pseudovectors ($\chi$), where $\chi=0.5\arctan (U/Q)$ and $U$ and $Q$ are the
Stokes parameters. The ring-like geometry of the polarization vectors at
large radii is centered approximately at the position of the submillimetre
source detected here, with the polarization fraction $P\sim20$\% at a radius
of 45\,kpc. Hayes et al. argue that such a polarization signal cannot be
generated {\it in situ}, but is more likely to be scattered light originating
from a central source. 

Our new observations, confirming the presence of a luminous dusty source
at the centre of the polarized emission supports the picture where Ly$\alpha$
photons escaping from a star-forming galaxy or AGN undergo a long flight,
before encountering H\,{\i} clouds in the CGM and scattering. As Hayes et al.
explain, it is the rare wing scatterings (i.e.\ photons with large $\Delta
\nu$ from the Ly$\alpha$ line center in the rest frame of an H\,{\sc i} cloud)
that (a) see lower optical depth and are more likely to escape the neutral
gas-filled halo and (b) have a higher probability of becoming polarized, with
a polarization fraction up to $P\sim40$\%. For these reasons, neither shock
excitation, photo-ionization or collisional excitation during gravitational
cooling in the in-flowing streams can explain the polarization observations,
although this is not to say the total Ly$\alpha$ luminosity is not contributed
to in part by non-scattering processes.

\subsection{A coherent model of SSA22-LAB1}

The detection of an SMG with $L_{\rm IR}\sim10^{12}{\rm L_\odot}$ located
between the lobes of R2 and R3 is a smoking gun: not only is the SMG spatially
coincident with the center of the polarization pattern, but its bolometric
luminosity, presumably dominated by far-infrared emission, indicates an ample
supply of Ly$\alpha$ photons, even if the majority of these are absorbed by
interstellar dust. Considering a naive model where the infrared luminosity is
produced entirely from re-processed starlight and represents the bulk of the
active star formation, we expect an {\it intrinsic} Ly$\alpha$ luminosity of
order $L^{\rm int}_{\rm Ly\alpha}\approx0.05L_{\rm IR}$ for Case B
recombination and a Salpeter initial mass function (Kennicutt et al.\ 1998,
Dijkstra \& Westra\ 2009). Thus, the escaping Ly$\alpha$ luminosity is $L^{\rm
esc}_{\rm Ly\alpha}\approx0.05f_{\rm esc}L_{\rm IR}$, where $f_{\rm esc}$ is
the escape fraction of Ly$\alpha$ photons from the central galaxy. However,
only a fraction, $f_{\rm scat}$, of these Ly$\alpha$ photons will go on to
scatter in the CGM into our line of sight and thus give rise to the LAB.

Three key factors affect $f_{\rm scat}$: (i) the cross-section for
Ly$\alpha$--H{\sc i} scattering for photons encountering surrounding neutral
gas; (ii) the overall filling factor of those clouds around the central source
and; (iii) the velocity distribution of the neutral gas determining how
photons can move out of resonance and free stream to the observer. Clearly (i)
is dependent on the $N_{\rm H}$ column, but since $f_{\rm scat}<1$ even for
filling factors of unity, this increases the escape fraction required from the
central source if scattered light is the sole source of extended Ly$\alpha$
emission. For $L_{\rm IR}=10^{12}{\rm L_\odot}$, this requires $f_{\rm
esc}>25$\% to account for the total $L_{\rm Ly\alpha}$ of R2 and R3. This is
not unreasonable if one considers a geometry where we are are viewing the
ULIRG system nearly edge-on, with a large optical depth to the star-forming
regions and thus any ultraviolet continuum emission from the central source
completely obscured along our line of sight but instead escaping in the plane
of the sky (see Geach et al.\ 2007). This geometry would also impart the
largest observable Ly$\alpha$ polarization signal, as photons escaping to our
line of sight would be scattered through 90$^\circ$, and would produce a
morphology consistent with the double lobed Ly$\alpha$ geometry hinted at by
the deep SAURON observations described in \S5.1.

\section{Conclusions}

\subsection{A model of LAB formation: a combination of `velvet rope' and
`bouncer' feedback?}

The observations imply that one of the main sources of Ly$\alpha$ photons
emitted by SSA22-LAB1 is either star formation or an AGN within the embedded
ULIRG. These photons are escaping the galaxy and scattering in cold gas distributed in
the CGM. Whatever the exact mechanism for Ly$\alpha$ emission, there must be a
large quantity of cold, neutral gas in the CGM. An open question therefore is
the origin of the circumgalactic gas responsible for scattering/emitting the
Ly$\alpha$ photons and giving rise to the observed extended emission-line
nebula. There are two options:

\begin{enumerate} \item The cold gas is comprised of pristine material
containing dense clumps of H{\sc i} being accreted onto the dark matter halo
(Zheng et al.\ 2011), potentially delivered by narrow streams within a hotter
medium. Ly$\alpha$ photons escaping the central galaxy illuminate this gas as
they scatter in the clumpy, neutral medium. The morphology of the resulting
LAB is therefore somewhat dependent on the geometry of the escaping photons
and the covering factor of the incoming clumps, which could be modified as the
inflow encounters radiation/outflows from the galaxy.

\item Neutral gas is being ejected into the CGM via a galactic outflow, having
already been accreted onto the galaxy. The outflow forms when shells of cold
gas are driven outwards by pressurized bubbles inflated within the ISM via
energy/momentum injection from stellar winds, supernovae and AGN. The outflow
strongly resembles a rarefaction wave escaping the galaxy along paths of least
resistance (e.g.,\ perpendicular to the disc), accelerating as they emerge into
the low-pressure CGM before fragmenting into cold clumps due to hydrodynamic
instabilities and eventually decelerating in the gravitational potential of
the halo (Dijkstra \& Kramer 2012, Creasey et al.\ 2013).\end{enumerate}

\noindent The key difference between (1) and (2) is the provenance and history
of the cold gas dominanting the CGM: pristine still-cooling or previously
cycled through a galaxy? The two scenarios describing the history of the CGM
can be related to the concept of `velvet rope' and `bouncer'
feedback\footnote{terms attributed to Neal Katz}. In velvet rope feedback, gas
never gets a chance to enter a galaxy where it can be assimilated into the
ISM. In contrast, bouncer mode describes feedback that ejects gas from the
galaxy back into the CGM.

Of course, these scenarios are not mutually exclusive. Gas must be delivered
to growing galaxies at some point, and this is likely to be close to, or
simultaneous with, the time the galaxies undergo their most rapid growth.
Therefore it is possible that emergent shells comprised of enriched gas could
also sweep-up cold, metal-free inflowing material that has not yet reached the
disc (thus, potentially preventing it from being accreted: velvet rope
feedback). In scenario 1, some of the inflowing gas might be blown back by an
emergent wind, and the observations suggest that the bulk of the line-emitting
gas is outflowing with respect to the central ULIRG (Fig.\ 2). One problem
with this scenario is that the covering factors of the streams are expected to
be very small ($f\sim1$--$3$ per cent; Faucher-Gigu\`ere \& Kere\v{s}\
2011). Outflows are likely to emerge preferentially between these streams,
allowing the cold inflowing material to continue to fuel the disc unarrested,
while cold interstellar material is simultaneously ejected (bouncer feedback).
However, the interaction of multi-phase galactic outflows and the narrow
cooling streams predicted by models is poorly understood, the hydrodynamic
interaction between the two is likely to be complicated, dependent on halo
mass, geometry and radius. Moreover, the halo will still have a diffuse
atmosphere, even if most of the mass is accreted along filamentary structure
(e.g.,\ Crain et al.\ 2013).

While the cold flows model predicts that the inflowing cold gas should also be
a source of Ly$\alpha$ radiation from collisional excitation, the polarization
observations show that a significant fraction of the Ly$\alpha$ emission is
not produced {\it in situ}. However, in a scenario where Ly$\alpha$ photons
from the central source are scattering in H{\sc i} clumps in these streams,
one must consider the low covering factors predicted for the inflowing cold
filaments. If the escaping Ly$\alpha$ photons are uncorrelated with the
geometry of the filaments (which could be a naive assumption if one considers
that the escape must preferentially be in the directions of lowest density,
i.e.,\ potentially {\it anti}-correlated with the filaments), then one would
expect a fraction $(1-f)$ of the Ly$\alpha$ photons {\it not} to scatter. In
effect, this drastically increases the required Ly$\alpha$ luminosity of the
central source. In contrast, gas entrained in a galaxy-scale wind can have
covering factors approaching unity, which arguably would result in an observed
Ly$\alpha$ morphology similar to R2 and R3.

Francis et al.\ (2013) argue that the CGM of Ly$\alpha$ Blob B1 ($z=2.38$)
could also be illuminated by scattered Ly$\alpha$ radiation, either
originating from a central source, or via fast shocks in the CGM. In the superwind
scenario, the fast shock is formed when hot gas driven by the emergent wind
from the central source collides with cold gas in the CGM. The detection of
C~{\sc iv} in the CGM of B1 is taken to be evidence for gas heating via fast
shocks (Francis et al.\ 2001), and Francis et al.\ (2013) argue that the
presence of carbon rules out a primordial origin for this material. Enriched
cold gas in the vicinity of low-mass satellites within the halo could be the
origin (these could also contribute Ly$\alpha$ luminosity via ultraviolet
leakage), as could metal-rich gas previously ejected from the central source.
Still, a primordial origin for at least a fraction of the cold gas cannot be
ruled out, and indeed should be expected.

\subsection{Final comments}

It seems that a coherent picture for the SSA22-LAB1 system is at last
emerging, although puzzles still remain. The SSA22-LAB1 system is clearly a
group in the early stage of formation, and while the Ly$\alpha$ `bridge'
linking the northern LBG C11 with the main emission structure may well be a
remnant cooling flow, the evidence suggests that a significant fraction of the
Ly$\alpha$ emission originates in the central ULIRG and scatters in H{\sc i}
clouds distributed in the CGM. The twist is that the central source does not
emit Ly$\alpha$ photons isotropically, indeed, we do not see ionizing photons
from the source, let alone the escaping Ly$\alpha$ radiation. Nevertheless,
the new SCUBA-2 observations presented here allow us to observe the
ultraviolet photons that are reprocessed by internal dust along our
line-of-sight. Although we do not see the ultraviolet continuum source
directly, abundant Ly$\alpha$ photons (and presumably a strong flux of
ionizing photons) must escape along an axis that is almost perpendicular to
the line of sight. Viewed along the emergent axis from another part of the
Universe we would almost certainly identify the source as a Ly$\alpha$ halo
containing an ultraviolet continuum source (as other LABs often do,
e.g.,\ Matsuda et al.\ 2004).

SSA22-LAB1 offers a fascinating insight into the physics of massive galaxy
formation: many of the ingredients of our latest models appear to be in action
in this one system, and it clearly provides a unique opportunity to study the
astrophysics of the galaxy/CGM interface. There are three key observations
that would provide more important clues:

\begin{enumerate}

 \item{Sensitive, high-resolution Ly$\alpha$ imaging. This would more clearly
reveal the morphology of the cold circumgalactic gas as traced by Ly$\alpha$,
providing the finer detail that might help distinguish an outflow with large
covering factor from narrow filamentary streams. SSA22-LAB1 is an ideal target
for the Multi Unit Spectroscopic Explorer (MUSE) IFU being commissioned on the
Very Large Telescope.}

 \item{Morphology of the dust continuum. High resolution sub-mm imaging would
more accurately pin-point the ULIRG within the Ly$\alpha$ halo and potentially
provide a clearer picture of the nature of the central luminous source,
including the geometry of the obscuring material. The Atacama Large Millimetre
Array (ALMA) is an obvious facility to perform this observation.}

 \item{Systemic redshift of the ULIRG. If the galaxy is responsible for a
biconical outflow in the plane of the sky (as suggested by the relative
velocities of the regions R2 and R3), then it should be very close in redshift
space to the Ly$\alpha$ peaks of R2 and R3. The ULIRG systemic redshift could
be measured with millimetre observations of high-{\it J} CO lines, which
should trace the densest molecular gas at the sites of star formation within
the ULIRG, and thus provide an accurate measurement of the systemic redshift
relative to the extended Ly$\alpha$. Again, this question could be addressed
by ALMA.} \end{enumerate}

\section*{Acknowledgements}

We are grateful to the anonymous referee, whose comments have improved the
clarity of this paper. We thank Matthew Hayes for providing the polarization
data shown in Figure 1, and for helpful discussions, Anne-Marie Weijmans for
providing the SAURON data cube and Rob Crain, Mark Dijkstra, and
Claude-Andr\'e Faucher-Gigu\`ere for useful discussions. J.E.G. acknowledges
support from the Royal Society by way of a University Research Fellowship.
J.S.D. acknowledges the support of the European Research Council via an
Advanced Grant, and the contribution of the EC FP7 SPACE project ASTRODEEP
(Ref.No: 312725). The James Clerk Maxwell Telescope is operated by the Joint
Astronomy center on behalf of the Science and Technology Facilities Council of
the United Kingdom, the National Research Council of Canada, and (until 31
March 2013) the Netherlands Organisation for Scientific Research. Additional
funds for the construction of SCUBA-2 were provided by the Canada Foundation
for Innovation.

\label{lastpage}

\end{document}